\documentclass[a4paper]{preamble/template}
\usepackage[utf8]{inputenc}

\usepackage{stix2}
\usepackage{amsmath}
\usepackage{graphicx}
\usepackage{epstopdf}
\usepackage{bm}
\usepackage{booktabs}
\usepackage{caption}
\usepackage{subcaption}
\usepackage{framed}
\usepackage{hyperref}
\usepackage{placeins}
\usepackage{tabularx}
\usepackage{multirow}

\newcommand{\vt}[1]{\mathbf{#1}}
\newcommand{\mat}[1]{\mathbf{#1}}
\newcommand{\transpose}{\mathsf{T}}
\newcommand{\euler}{\mathrm{e}}

\graphicspath{{./figures/}}

\heading{Syver Døving Agdestein and Benjamin Sanderse}

\title{
    Learning filtered discretization operators: non-intrusive versus intrusive approaches
}

\author{SYVER DØVING AGDESTEIN$^{1*}$ AND BENJAMIN SANDERSE$^{2*}$}

\address{
    $^{1}$ e-mail: syver.agdestein@cwi.nl, https://www.cwi.nl/people/syver-agdestein
    \and
    $^{2}$ e-mail: b.sanderse@cwi.nl, https://www.cwi.nl/$\sim$sanderse
    \and
    $^{*}$ Centrum Wiskunde \& Informatica (CWI) \\
    Science Park 123, 1098XK Amsterdam, Netherlands
}


\abstract{
    Simulating multi-scale phenomena such as turbulent fluid flows is typically
    computationally very expensive. Filtering the smaller scales allows for using coarse
    discretizations, however, this requires closure models to account for the effects of the
    unresolved on the resolved scales. The common approach is to filter the continuous
    equations, but this gives rise to several commutator errors due to nonlinear terms,
    non-uniform filters, or boundary conditions.

    We propose a new approach to filtering, where the equations are discretized first and
    then filtered. For a non-uniform filter applied to the linear convection equation, we
    show that the discretely filtered convection operator can be inferred using three
    methods: intrusive (`explicit reconstruction')  or non-intrusive operator inference,
    either via `derivative fitting' or `trajectory fitting' (embedded learning). We show
    that explicit reconstruction and derivative fitting identify a similar operator and
    produce small errors, but that trajectory fitting requires significant effort to train
    to achieve similar performance. However, the explicit reconstruction approach is more
    prone to instabilities. 
}

\begin{document}

\maketitle

\section{INTRODUCTION}

Turbulent fluid flows are characterized by the presence of high frequencies. Direct
numerical simulation (DNS) aims to resolve all turbulent frequencies at the expense of a
high computational cost. In contrast, approaches such as large eddy simulation (LES) consist
of modelling the behavior of the large features only, i.e.\ the low frequency components.
Mathematically, this implies the use of low-pass filters to obtain equations for the low
frequency components. These filters do typically not commute with the differential
operators, nonlinear terms, or boundary conditions \cite{VanDerVen1995}. Furthermore,
requiring that they do commute imposes severe restrictions on the choice of filters, e.g.\
by requiring vanishing moments or spatial uniformity
\cite{VanDerVen1995,Vasilyev1998,Marsden2002}. This often conflicts with other restrictions,
such as requiring the filter to be linked to or directly deduced from the spatial
discretization (\emph{implicit} LES) \cite{Boris1990,Boris1992}.

Traditionally, the equations are filtered on the continuous level, and then discretized. The
commutator errors arising from the filter are then modelled in terms of the filtered
solution only, through a closure model \cite{Sagaut2006}. Recent efforts have been made to
learn the closure model \emph{for a given filter and discretization} using deep neural
networks \cite{Bar-Sinai2019,Beck2021}, thus closing the equations after discretization.

In this article, we continue on this discrete approach, in which we aim to find discrete
representations of filtered operators, constructed by using high-fidelity simulations. We
compare three different approaches: (i) explicit construction of the filtered operator using
quadrature and reconstruction operators; (ii) least-squares reconstruction of the filtered
operator, analogous to operator inference; (iii) reconstruction of the embedded filtered
operator, analogous to ``solver in the loop'' methods \cite{Kochkov2021,List2022,Um2020}.

We focus specifically on commutator errors arising from non-uniform filtering of linear
operators (in which case filtering and differentiation do not commute), so that the filtered
operators that need to be constructed remain linear, and we do not need to use complex
function approximators such as neural networks. This allows us to get a clean picture of how
these three data-driven methods perform in constructing a filtered operator. In Section
\ref{sec:problem-statement}, we present the discrete problem. In Section
\ref{sec:operator-inference}, we present the three approaches for building the discrete
filtered operator. In Section \ref{sec:results}, we show the results for the filtered
convection equation, comparing the three approaches for two different filters.

\section{PROBLEM STATEMENT} \label{sec:problem-statement}

In this paper we consider the one-dimensional linear partial differential equation
\begin{equation} \label{eqn:continuous}
    \frac{\partial u}{\partial t}(x, t) = \mathcal{A} u(x, t), \quad x \in \Omega, \quad t >
    0,
\end{equation}
where $\Omega = [0, 1]$ is a periodic domain, $\mathcal{A}$ is a linear differential
operator with respect to the spatial variable $x$, and $u(x, 0) = u_0(x)$ are the initial
conditions.

A uniform periodic discretization of $\Omega$ is given by $\vt{x}^{(N)} = (x_n^{(N)})_{n =
1}^N \in \mathbb{R}^N$, where $N$ is the number of grid points, $x_n^{(N)} = n \Delta
x^{(N)}$, and $\Delta x^{(N)} = \frac{1}{N}$. A semi-discrete approximation to
\eqref{eqn:continuous} is then given by
\begin{equation} \label{eqn:discrete}
    \frac{\mathrm{d} \vt{u}}{\mathrm{d} t}(t) = \mat{A} \vt{u}(t), \quad t > 0,
\end{equation}
where $\vt{u}(t) \in \mathbb{R}^N$ is a semi-discrete solution and $\mat{A} \in
\mathbb{R}^{N \times N}$ is a discrete approximation of the differential operator
$\mathcal{A}$, taking into account the periodicity. The exact semi-discrete solution is
given by $\vt{u}(t) = \euler^{t \mat{A}} \vt{u}_0$.

We are interested in the case where the spectral content of $u$ is too large to be
represented by the discrete solution. A low-pass filter is thus applied to obtain a new set
of equations that can be resolved with a reasonably small $N$. We here consider a general
linear continuous kernel filter $\mathcal{F}$ defined by
\begin{equation} \label{eqn:filter}
    \bar{u}(x) = \mathcal{F}(u)(x)
    = \int_\mathbb{R} G(x, \xi) u(\xi) \, \mathrm{d} \xi
    % = \sum_{z \in \mathbb{Z}} \int_\Omega G(x, \xi + z) u(\xi) \, \mathrm{d} \xi
    \quad \forall x \in \Omega,
\end{equation}
where $\bar{u}$ is the filtered solution and $G : \Omega \times \mathbb{R} \to \mathbb{R}$
is the kernel. Note that the function $u$ is extended by periodicity to $\mathbb{R}$, i.e.
$u(x) = u(x \mod |\Omega|)$ for $x \not\in \Omega$, thus allowing for infinitely long filter
kernels such as a Gaussian kernel. The filter is non-uniform in space, so it is in general
not of convolution type (in which case one would have $G(x, \xi) = G(\xi - x) \ \forall (x,
\xi)$).

In many closure modelling approaches, the equations are filtered on the continuous level:
\begin{equation} \label{eqn:continuous_filtered}
    \frac{\partial \bar{u}}{\partial t}
    % = \mathcal{F}(\mathcal{A} u)
    = \overline{(\mathcal{A} u)}
    = \mathcal{A} \bar{u} + \mathcal{C}(u, \bar{u}),
\end{equation}
where $\mathcal{C}(u, \bar{u}) = \overline{(\mathcal{A} u)} - \mathcal{A} \bar{u}$ is the
commutator error, typically caused by nonlinear term. In the current setting where
$\mathcal{A}$ is linear, a commutator error still arises due to the fact that non-uniform
filtering and differentiation do not commute. The above equations are \emph{unclosed} in the
sense that $\mathcal{C}$ depends on $u$, and typically has to be modelled as a function of
$\bar{u}$ only, i.e.\ $\mathcal{C}(u, \bar{u}) \approx \mathcal{M}(\bar{u})$, where
$\mathcal{M}$ is a \emph{closure model}, for example obtained by truncating the continuous
Taylor series expansion of $u$ in the integral \eqref{eqn:filter} \cite{Ghosal1995}. The
filtered equations are then discretized on a coarse grid $\vt{x}^{(M)}$, $M \leq N$:
\begin{equation} \label{eqn:filtered_then_discretized}
    \frac{\mathrm{d} \bar{\vt{u}}}{\mathrm{d} t}(t) = \mat{A}^{(M)} \bar{\vt{u}}(t) +
    \vt{M}(\bar{\vt{u}}(t)), \quad t > 0,
\end{equation}
where $\bar{\vt{u}}(t) \in \mathbb{R}^M$ is an approximation to $\bar{u}(\vt{x}^{(M)}, t)$,
and $\mat{A}^{(M)} \in \mathbb{R}^{M \times M}$ and $\vt{M} : \mathbb{R}^M \to \mathbb{R}^M$
are coarse-grid approximations of the operators $\mathcal{A}$ and $\mathcal{M}$
respectively. In this approach one thus first commits a continuous closure model error, and
then a (coarse scale) discretization error.

We propose to obtain the discrete equations for the filtered solution in a different way, by
\textit{discretizing first}, and \textit{then filtering}. We employ a discrete approximation
of the filter, represented by a matrix $\mat{W}$ built using a \emph{quadrature rule}:
\begin{equation}
     \bar{\vt{u}} = \mat{W} \vt{u}.
\end{equation}
Note that the filtered and unfiltered discrete solutions $\bar{\vt{u}}$ and $\vt{u}$ are
represented on different grids $\vt{x}^{(M)}$ and $\vt{x}^{(N)}$, with $\bar{\vt{u}}(t) \in
\mathbb{R}^M$ and $\vt{u}(t) \in \mathbb{R}^N$, $M \leq N$. An accurate discrete filter
should be such that $\sum_n W_{m n} u(x_n^{(N)}, t) \approx \bar{u}(x_m^{(M)}, t)$, $m \in
\{1, \dots, M\}$. It is assumed that all the relevant frequencies of $u$ are resolved on
$\vt{x}^{(N)}$, but not necessarily on $\vt{x}^{(M)}$. To close the filtered equations, the
`discretize first' approach  requires an approximation of the inverse filtering operator by
using a reconstruction rule, represented by a matrix $\mat{R} \in \mathbb{R}^{N \times M}$,
such that $\vt{u} \approx \mat{R} \bar{\vt{u}}$. The discretely filtered solution
$\bar{\vt{u}}$ is then obtained by solving the equations
\begin{equation} \label{eqn:discrete_filtered}
    \frac{\mathrm{d} \bar{\vt{u}}}{\mathrm{d} t}(t) = \bar{\mat{A}} \bar{\vt{u}}(t), \quad
    t > 0,
\end{equation}
where $\bar{\mat{A}} = \mat{W} \mat{A} \mat{R}$ is the discretely filtered differential
operator and the filtered initial conditions are given by $\bar{\vt{u}}(0) = \mat{W}
\vt{u}_0$. Since the unfiltered solution is reconstructed directly, Equation
\eqref{eqn:discrete_filtered} is closed. In this approach one thus commits a (fine-grid)
discretization error, a quadrature error, and a reconstruction error. The choice of
$\bar{\mat{A}}$ may be seen as an effective closure model, and the resulting coarse-grid
commutator error $\bar{\mat{A}} - \mat{A}^{(M)}$ can be computed, although it is not
explicitly modelled (nor is it needed).

The problem that we address in this article is now the following: given a fine-scale
discretization $\mat{A}$ and a quadrature rule $\mat{W}$, can we infer $\bar{\mat{A}}$? To
answer this question, we consider two different approaches, that are outlined in the next
section.

\section{DATA-DRIVEN OPERATOR INFERENCE} \label{sec:operator-inference}

The two data-driven approaches to infer the matrix $\bar{\mat{A}}$ considered in this
article are:
\begin{itemize}
    \item \textit{Indirectly} building $\bar{\mat{A}}$ by first explicitly building
        $\mat{W}$, $\mat{A}$, and $\mat{R}$ and then computing  $\bar{\mat{A}} = \mat{W}
        \mat{A} \mat{R}$, see section \ref{sec:intrusive}. This approach is
        \textit{intrusive} in the sense that $\mat{A} \in \mathbb{R}^{N \times N}$ is
        required.
    \item \textit{Directly} building the reduced order operator $\bar{\mat{A}} \in
        \mathbb{R}^{M \times M}$ (see section \ref{sec:implicit_reconstruction}), using
        gradient-free derivative fitting (\ref{sec:derivative-fit}) and gradient-based
        embedded (\ref{sec:embedded}) approaches. This approach is \textit{non-intrusive}
        since access to $\mat{A}$ is not required.
\end{itemize}

In the following, we will employ the snapshot matrices $\mat{U} \in \mathbb{R}^{N \times
d}$, $\dot{\mat{U}} \in \mathbb{R}^{N \times d}$, $\bar{\mat{U}} \in \mathbb{R}^{M \times
d}$, and $\dot{\bar{\mat{U}}} \in \mathbb{R}^{M \times d}$, containing unfiltered and
filtered discrete solutions and their time derivatives at different time instances. Here, $d
= n_\text{IC} n_t$ is the number of snapshots, $n_\text{IC}$ is the number of initial
conditions, and $n_t$ is the number of time instances at which the solutions are evaluated.

\subsection{Indirect approach: building the reconstruction matrix $\mat{R}$} \label{sec:intrusive}

Using the midpoint rule \cite{Hansen2010}, the quadrature weights $\mat{W}$ used for
approximating the weighted integral \eqref{eqn:filter} are constructed as follows:
\begin{equation} \label{eqn:quadrature}
    W_{m n} = \frac{\tilde{G}(x_m, \xi_n)}{\sum_{i = 1}^N \tilde{G}(x_m, \xi_i)},
\end{equation}
where $\tilde{G}(x, \xi) = \sum_{z \in \mathbb{Z}} G(x, \xi + z |\Omega|)$ accounts for the
periodicity of $\Omega$ and potentially infinitely long kernel $G$. The sum is truncated to
$z \in \{-1, 0, 1\}$ in the case of a sufficiently local filter support (e.g.\ for the
Gaussian kernel). The normalization factor ensures that constant functions are preserved
upon filtering. Note that higher order quadrature rules may be built by requiring that
certain classes of functions are filtered exactly, but the simple rule
\eqref{eqn:quadrature} is sufficient for our purposes (since $N$ is assumed to be large) and
has the advantage that it leads to positive weights if the kernel is positive.

Finding the reconstruction matrix $\mat{R}$ is not straightforward, as the matrix $\mat{W}$
is in general not invertible ($M \neq N$). $\mat{R}$ is therefore built by minimizing the
expectation of the reconstruction error for a certain class of functions $\mathcal{U}$:
\begin{equation}
    \underset{\mat{R} \in \mathbb{R}^{N \times M}}{\min} \mathbb{E}_{u \sim
    \mathcal{U}} L^\text{fit}(\mat{R}, \bar{u}(\vt{x}^{(M)}), u(\vt{x}^{(N)})) +
    \lambda L^\text{prior}(\mat{R}),
\end{equation}
where $L^\text{fit}$ measures the reconstruction error, $L^\text{prior}$ penalizes deviation
from prior assumptions on the form of $\mat{R}$, and $\lambda$ is a regularization
parameter. Setting the accuracy metric as the mean squared error
\begin{equation}
    L^\text{fit}(\mat{R}, \bar{\vt{u}}, \vt{u}) = \| \mat{R} \bar{\vt{u}} - \vt{u} \|^2
\end{equation}
leads to a class of least squares problems. We consider the case where
$L^\text{prior}(\mat{R}) = \|\mat{R}\|_F^2 = \sum_{n m} R_{n m}^2$, and after approximating
the expectation value by a mean over the $d$ snapshots $\mat{U}$ and $\bar{\mat{U}}$, we get
\begin{equation} \label{eqn:R_data_driven}
    \mat{R} = \underset{\mat{R} \in \mathbb{R}^{N \times M}}{\operatorname{argmin}} \,
    \frac{1}{d} \| \mat{R} \bar{\mat{U}} - \mat{U} \|_F^2 + \lambda \| \mat{R} \|_F^2 =
    \mat{U} \bar{\mat{U}}^\transpose (\bar{\mat{U}} \bar{\mat{U}}^\transpose + d \lambda
    \mat{I})^{-1}.
\end{equation}
The resulting operator $\bar{\mat{A}}^\text{int} = \mat{W} \mat{A} \mat{R}$ is labelled as
\emph{intrusive} since it requires access to the full order model operator $\mat{A}$.

\subsection{Direct approach: building the filtered operator $\bar{\mat{A}}$}
\label{sec:implicit_reconstruction}

The operator $\bar{\mat{A}}$ can also be constructed non-intrusively from filtered data. The
solver used for \eqref{eqn:discrete} is then considered to be a ``black box'', which returns
data samples of the solution and its time derivative. Filtered samples are obtained by
applying the discrete filter $\mat{W}$. The operator $\bar{\mat{A}}$ then follows from a
minimization problem of the form
\begin{equation}
    \underset{\bar{\mat{A}} \in \mathbb{R}^{M \times M}}{\min} \mathbb{E}_{\substack{u \sim
    \mathcal{U} \\ t \sim \mathcal{T}}} L^\text{fit}(\bar{\mat{A}}, \bar{u}(\vt{x}^{(M)},
    \cdot), t) + \lambda_\text{prior} L^\text{prior}(\bar{\mat{A}}) + \lambda_\text{stab}
    L^\text{stab}(\bar{\mat{A}}),
\end{equation}
where $\mathcal{U}$ is the space of solutions to \eqref{eqn:continuous}, $\mathcal{T}$ is a
distribution of expected time instances, for example a uniform distribution on $[0, T]$ for
some time horizon $T$, $L^\text{prior}$ penalizes deviation from prior expected properties,
and $L^\text{stab}$ penalizes numerical instabilities that arise from the non-uniform filter
(namely anti-diffusion caused by the non-uniformity of the filter). In the following, we set
$L^\text{stab}(\bar{\mat{A}}) = \| \bar{\mat{A}} - \mat{D}^{(M)} \|_F^2$, where
$\mat{D}^{(M)}$ is a discretization of the diffusion operator $\frac{\partial^2}{\partial
x^2}$ on $\vt{x}^{(M)}$. This limits anti-diffusion due to eigenvalues with positive real
part. In addition, we set $L^\text{prior}(\bar{\mat{A}}) = \| \bar{\mat{A}} - \mat{A}^{(M)}
\|_F^2$, motivated by the assumption that the behavior of the filtered solution should be
similar to the one of the unfiltered solution (as is the reason for doing LES in the first
place), resulting in a commutator error that is small compared to the exact right-hand size,
i.e.\ $\mathcal{O} \left( \frac{\| \bar{\mat{A}} - \mat{A}^{(M)} \|_F}{\|\mat{A}^{(M)}\|_F}
\right) = \mathcal{O} \left( \frac{\|\mathcal{C}\|}{\|\mathcal{A}\|} \right) \ll 1$.

$L^\text{fit}$ is an instantaneous metric evaluating the performance of the operator
$\bar{\mat{A}}$ with respect to a discrete filtered solution $\bar{\vt{u}}$, for which two
different choices will be proposed, as explained in the next two sections.

\subsubsection{Gradient-free operator inference (derivative fitting)} \label{sec:derivative-fit}

The first choice for $L^\text{fit}$ is similar to the operator inference framework presented
in \cite{Peherstorfer2016b}, where the authors propose to fit the operators describing an
ODE to data samples of the time derivative, by using performance metrics of the form
\begin{equation}
    L^\text{fit}(\bar{\mat{A}}, \bar{\vt{u}}, t) = \left \| \bar{\mat{A}} \bar{\vt{u}} -
    \frac{\mathrm{d} \bar{\vt{u}}}{\mathrm{d} t} \right \|^2(t).
\end{equation}
We call this approach `derivative fitting'. Note that with this performance metric, neither
the filtered solution $\bar{\vt{u}}$ nor its time derivative $\frac{\mathrm{d}
\bar{\vt{u}}}{\mathrm{d} t}$ depend on $\bar{\mat{A}}$, and that the dependency on
$\bar{\mat{A}}$ is quadratic in the loss function. As in Equation \eqref{eqn:R_data_driven},
approximating the expectation value by a sum over the $d$ snapshots results in the following
closed form expression for the operator:
\begin{equation} \label{eqn:derivative_fit}
    \begin{split}
        \bar{\mat{A}}^\text{DF} & = \underset{\bar{\mat{A}} \in \mathbb{R}^{M \times
        M}}{\operatorname{argmin}} \, \frac{1}{d} \| \bar{\mat{A}} \bar{\mat{U}} -
        \dot{\bar{\mat{U}}} \|_F^2 + \lambda_\text{prior} \| \bar{\mat{A}} - \mat{A}^{(M)}
        \|_F^2 + \lambda_\text{stab} \| \bar{\mat{A}} - \mat{D}^{(M)} \|_F^2 \\
        & = \left(\frac{1}{d} \dot{\bar{\mat{U}}} \bar{\mat{U}}^\transpose +
        \lambda_\text{prior} \mat{A}^{(M)} + \lambda_\text{stab} \mat{D}^{(M)}\right)
        \left(\frac{1}{d} \bar{\mat{U}} \bar{\mat{U}}^\transpose + (\lambda_\text{prior} +
        \lambda_\text{stab}) \mat{I}\right)^{-1},
    \end{split}
\end{equation}
where DF denotes derivative fitting.

\subsubsection{Embedded operator inference (solver-in-the-loop)} \label{sec:embedded}

The second choice for $L^\text{fit}$ is the instantaneous performance metric
\begin{equation}
    L^\text{fit}(\bar{\mat{A}}, \bar{\vt{u}}, t) = \| \vt{S}(\bar{\mat{A}}, \bar{\vt{u}}(0),
    t) - \bar{\vt{u}}(t) \|^2.
\end{equation}
In this expression we introduced the notion of the ODE solver $\vt{S} : \mathbb{R}^{M \times
M} \times \mathbb{R}^M \times \mathbb{R} \to \mathbb{R}^{M}$, where $\vt{S}(\bar{\mat{A}},
\bar{\vt{u}}_0, t)$ approximates the solution to \eqref{eqn:discrete_filtered} for the given
operator $\bar{\mat{A}}$ and filtered initial conditions $\bar{\vt{u}}_0$. In the previous
section, the operator $\bar{\mat{A}}^\text{DF}$ was obtained by fitting known filtered
solution trajectories $\bar{\vt{u}}$ to their time derivatives $\frac{\mathrm{d}
\bar{\vt{u}}}{\mathrm{d} t}$ using least squares. In contrast, in this section the operator
$\bar{\mat{A}}$ is inferred by fitting the predicted solution trajectories
$\vt{S}(\bar{\mat{A}}, \bar{\vt{u}}(0), \cdot)$ to the known filtered solution trajectories
$\bar{\vt{u}}$. The operator $\bar{\mat{A}}$ is thus \emph{embedded} in the solver $\vt{S}$
\cite{Kochkov2021,List2022,Um2020}.

Assuming the program $\vt{S}$ is differentiable, i.e.\ we have access to the quantity
$\frac{\partial \vt{S}}{\partial \bar{\mat{A}}}(\bar{\mat{A}}, \bar{\vt{u}}_0, t) \in
\mathbb{R}^{M \times M \times M}$, we will use a gradient-descent based optimization
algorithm to identify the correct $\bar{\mat{A}}$, using adjoint-mode differentiation with
checkpointing to compute the vector-Jacobian products \cite{Chen2018,Rackauckas2020}.

Note that each evaluation of the loss function requires solving the ODE
\eqref{eqn:discrete_filtered} of size $M$ for the given operator $\bar{\mat{A}}$
(solver-in-the-loop). For more computationally expensive problems, e.g.\ three-dimensional
non-linear equations such as the Navier-Stokes equations, evaluating $L^\text{fit}$ at all
initial conditions and time steps in the training dataset can quickly get expensive. We will
use \emph{stochastic} gradient descent \cite{Bottou1991} to reduce this computational cost,
by evaluating the loss and its gradient for only a few initial conditions and time instances
at a time. In the following, the operator fitted while embedded in the solver will be
denoted by $\bar{\mat{A}}^\text{emb}$.

\section{NUMERICAL RESULTS: CONVECTION EQUATION} \label{sec:results}

We consider the convection equation with unit velocity, obtained by setting $\mathcal{A} =
-\frac{\partial}{\partial x}$ in \eqref{eqn:continuous}. To test the effectiveness of the
three approaches to infer the filtered operator $\bar{\mat{A}}$, we consider two filters:
the top-hat filter, for which
\begin{equation} \label{eqn:top_hat}
    G(x, \xi) = \begin{cases}
        \frac{1}{2 h(x)}, \quad & |\xi - x| \leq h(x), \\
        0, \quad & \text{otherwise},
    \end{cases}
\end{equation}
and the Gaussian filter, for which
\begin{equation} \label{eqn:gaussian}
    G(x, \xi) = \sqrt{\frac{3}{2 \pi h^2(x)}} \euler^{-\frac{3 (\xi - x)^2}{2 h^2(x)}},
\end{equation}
where $h(x)$ is a variable filter radius. For both filters, it is chosen to be
\begin{equation} \label{eqn:filter_width}
    h(x) = \left( 1 + \frac{1}{3} \sin(2 \pi x) \right) h_0
\end{equation}
with $h_0 = \frac{1}{50}$. At $x = \frac{1}{4}$, the filter radius is thus twice as large as
at $x = \frac{3}{4}$, and filtered waves moving from $x = \frac{1}{4}$ to $x = \frac{3}{4}$
are subject to \emph{anti-diffusion}, where high-frequency components grow. Small numerical
errors get amplified on this interval, and finding a stable $\bar{\mat{A}}$ is thus not
straightforward. Note also that the Gaussian standard deviation $\frac{h}{\sqrt{3}}$ is
chosen to give the same Taylor series expansion for the local attenuation factor as a
function of $k h$ in spectral space as for the top-hat filter, where $k$ is the frequency.

Given initial conditions, closed form expressions are available for the unfiltered and
filtered solutions and their time derivatives. However, we use DNS to generate data samples
(by solving Equation \eqref{eqn:discrete}), as this is the approach one would take for other
equations that do not have closed-form solutions. For this purpose, the continuous
convection operator $\mathcal{A}$ is discretized using a sixth order central difference
stencil on the fine grid $\vt{x}^{(N)}$:
\begin{equation}
    \mat{A} = \frac{1}{60 \Delta x^{(N)}} \operatorname{circ}_N(1, -9, 45, 0, -45, 9, -1),
\end{equation}
and similarly for $\mat{A}^{(M)}$, where $\operatorname{circ}_\eta(s_{-\alpha}, \dots,
s_{\alpha}) \in \mathbb{R}^{\eta \times \eta}$ denotes a circulant matrix where the $i$-th
periodically extended superdiagonal is filled with the stencil entry $s_i$ for $i \in
\{-\alpha, \dots, \alpha\}$. Similarly, we build $\mat{D}^{(M)} = \frac{1}{180 \left(\Delta
x^{(M)}\right)^2} \operatorname{circ}_M(2, -27, 270, -490, 270, -27, 2)$ (of order six).

The unfiltered DNS equations \eqref{eqn:discrete} are discretized with $N = 1000$ points,
and solved using an adaptive fourth order Runge-Kutta time integrator \cite{Tsitouras2011}
with absolute and relative tolerances $\epsilon_\text{abs} = 10^{-10}$ and
$\epsilon_\text{rel} = 10^{-8}$. The resulting solution is saved at $n_t$ time points. The
time derivative snapshots are obtained by evaluating the DNS right hand side $\mat{A}
\vt{u}$. The filters are discretized using the periodically extended quadrature rule
\eqref{eqn:quadrature}. The resulting $\mat{W}$ is then used to generate the reference
solutions $\mat{W} \vt{u}$ and time derivatives $\mat{W} \mat{A} \vt{u}$ for a given $M$.

\begin{figure}
    \centering
    \begin{tabular}{ccc}
        Filter radius $h(x)$ & Data sample $u(x)$ & Fourier coefficients $|c_k|$ \\
        \includegraphics[width=0.3\textwidth]{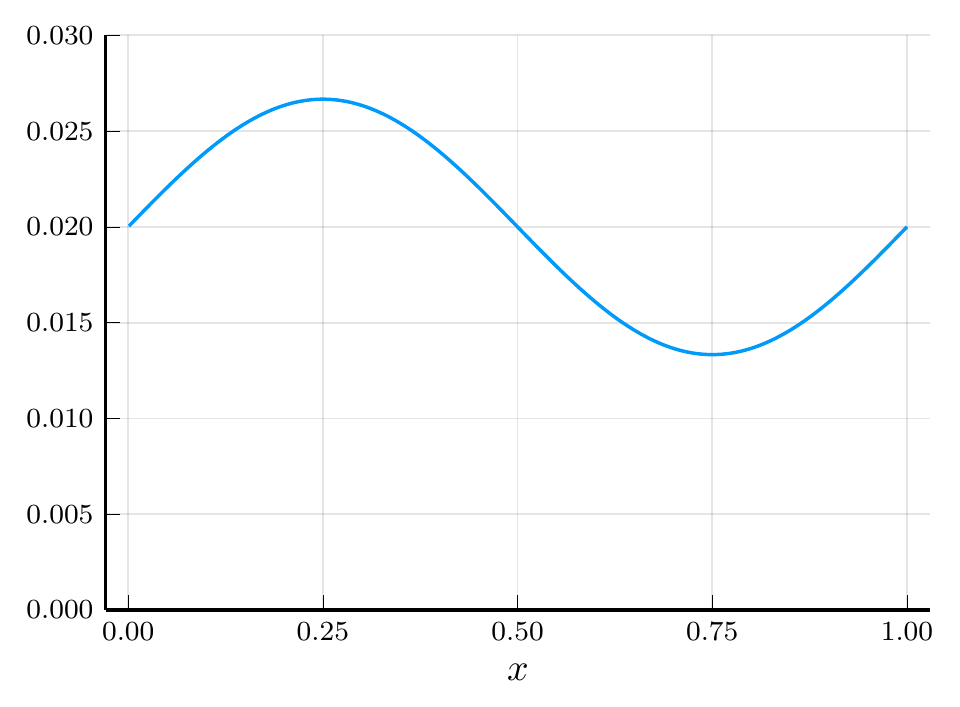} &
        \includegraphics[width=0.3\textwidth]{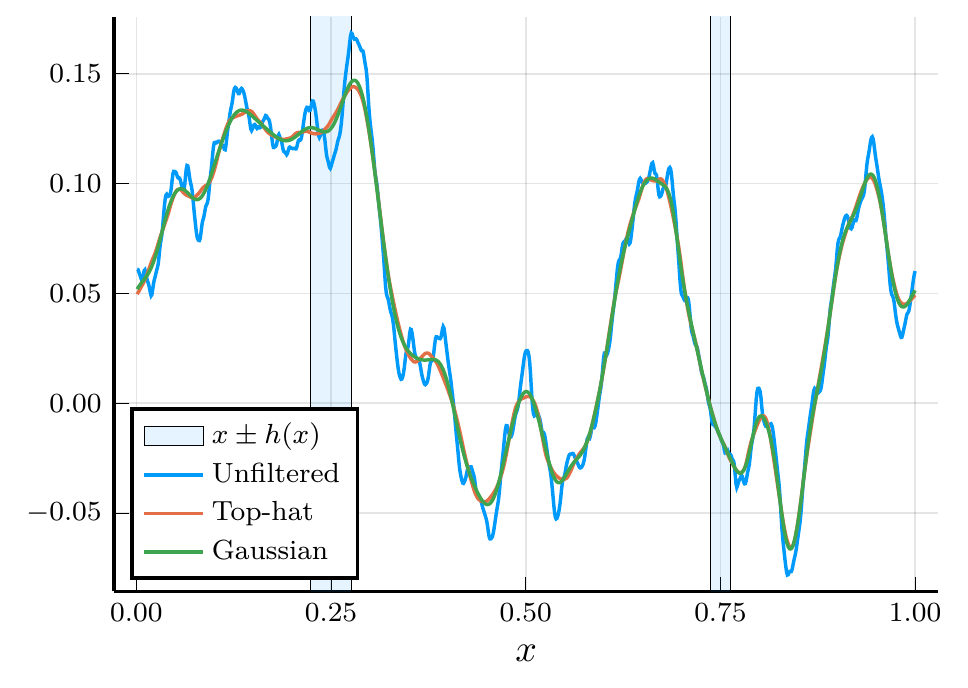} &
        \includegraphics[width=0.3\textwidth]{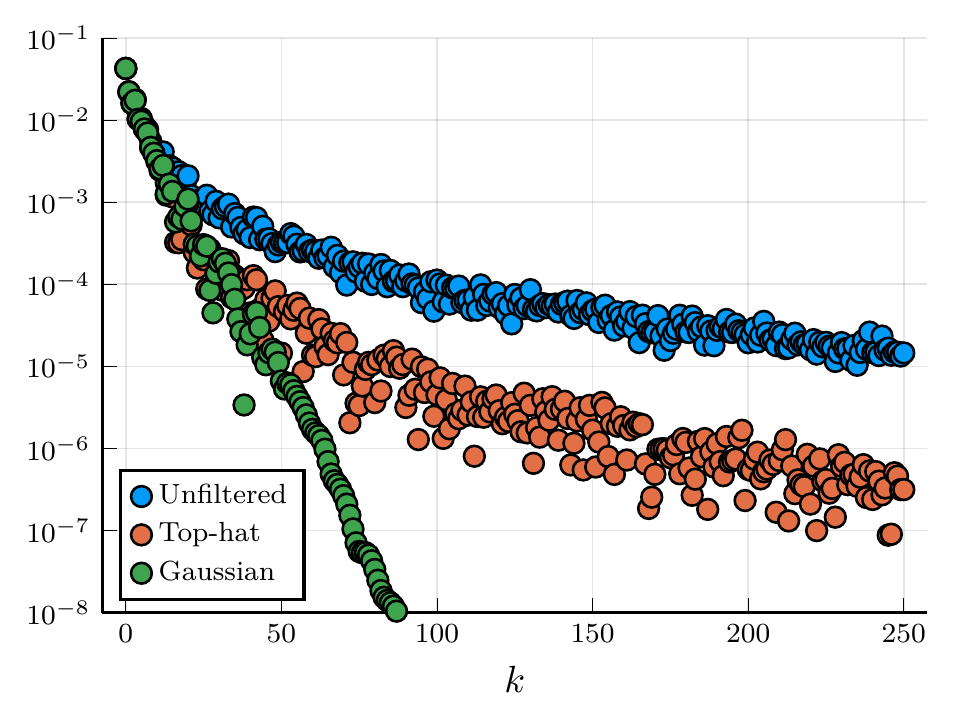} \\
    \end{tabular}
    \caption{
        Left: Filter radius $h$ used in the simulations. Middle and right: Unfiltered and
        filtered initial conditions and their Fourier coefficients for one of the training
        samples.
    }
    \label{fig:filters}
\end{figure}

Four datasets $\mathcal{D} \in \{\mathcal{D}^\text{train}, \mathcal{D}^\text{valid},
\mathcal{D}^\text{test}, \mathcal{D}^\text{long}\}$ are created. They consist of waves with
a maximum frequency of $K = 250$ and $n_\text{IC}^\mathcal{D}$ different initial conditions
$(u_0^{\mathcal{D}, i})_{i \in \{1, \dots, n_\text{IC}^\mathcal{D}\}}$ given by
\begin{equation}
    u_0^{\mathcal{D}, i}(x) = \sum_{k = 0}^K \frac{1 + \epsilon^{\mathcal{D}, i}}{(5 +
    k)^2} \cos(2 \pi k x + \theta^{\mathcal{D}, i}), \quad i \in \{1, \dots,
    n_\text{IC}^\mathcal{D}\},
\end{equation}
where $\epsilon^{\mathcal{D}, i} \sim \mathcal{N}(0, \frac{1}{5})$ and $\theta^{\mathcal{D},
i} \sim \mathcal{U}([0, 2 \pi])$ are sampled independently for each $\mathcal{D}$ and $i$.
Note that the fine grid $\vt{x}^{(N)}$ resolves all the $K$ frequencies. This choice of
coefficients creates a profile where the ``turbulent'' features (high frequencies) are small
compared to the large features, but are still visually different from their filtered
counterparts. One of the resulting initial conditions and its coefficients are shown for
both filters in Figure \ref{fig:filters}, along with the common filter radius $h$. 

The parameters for the datasets are given in Table \ref{tab:datasets}.
$\mathcal{D}^\text{train}$ is used to fit the operators. $\mathcal{D}^\text{valid}$ is used
to choose the hyperparameters $\lambda$. $\mathcal{D}^\text{test}$ is used to measure the
generalization capacity of the inferred models. $\mathcal{D}^\text{long}$ is used to measure
the long term stability of the operators.

\begin{table}
    \centering
    \begin{tabular}{|c|ccc|} \hline
        $\mathcal{D}$ & $n_\text{IC}$ & $n_t$ & $T$ \\ \hline
        train & $1000$ & $50$ & $0.1$ \\
        valid & $20$ & $10$ & $1.0$ \\
        test & $100$ & $20$ & $1.0$ \\
        long & $50$ & $500$ & $100.0$ \\ \hline
    \end{tabular}
    \caption{Parameters used to build the different datasets $\mathcal{D}$: number of
    initial conditions ($n_\text{IC}^\mathcal{D}$), number of time points
    ($n_t^\mathcal{D}$), and convection time ($T^\mathcal{D}$).}
    \label{tab:datasets}
\end{table}

\begin{figure}
    \begin{center}
        \renewcommand{\tabularxcolumn}[1]{m{#1}}
        \begin{tabularx}{\textwidth}{
            >{\centering\arraybackslash}m{0.02\textwidth}
            >{\centering\arraybackslash}X
            >{\centering\arraybackslash}X
            >{\centering\arraybackslash}X
            >{\centering\arraybackslash}X
            >{\centering\arraybackslash}X
        }
            % & $\mat{W}$
            & $\mat{R}$
            % & $\bar{\mat{A}}^\text{int}$
            % & $\bar{\mat{A}}^\text{emb}$ \\
            & $\bar{\mat{A}}^\text{int} - \mat{A}^{(M)}$
            & $\bar{\mat{A}}^\text{DF} - \mat{A}^{(M)}$
            & $\bar{\mat{A}}^\text{emb} - \mat{A}^{(M)}$ \\
            \rotatebox[origin=c]{90}{Top-hat} &
            \includegraphics[width=0.22\textwidth]{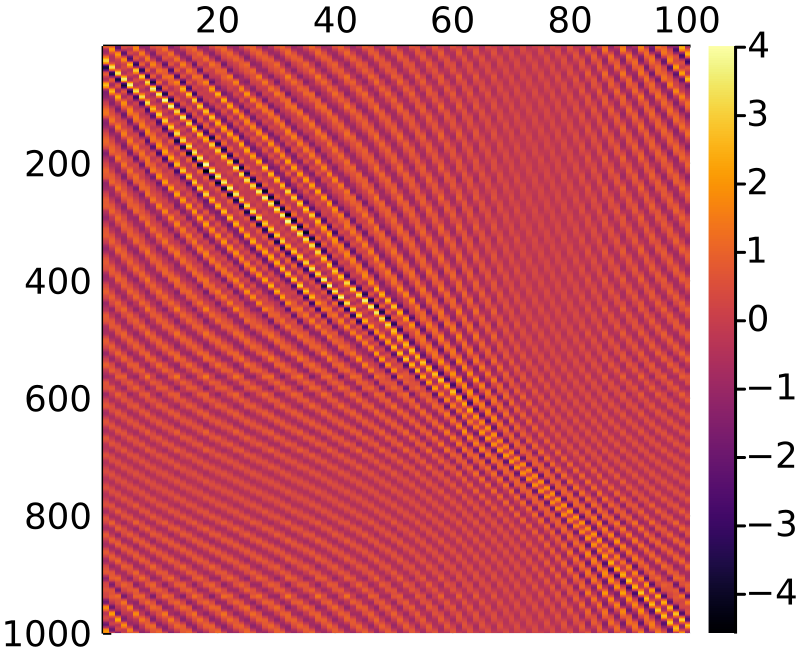} &
            \includegraphics[width=0.22\textwidth]{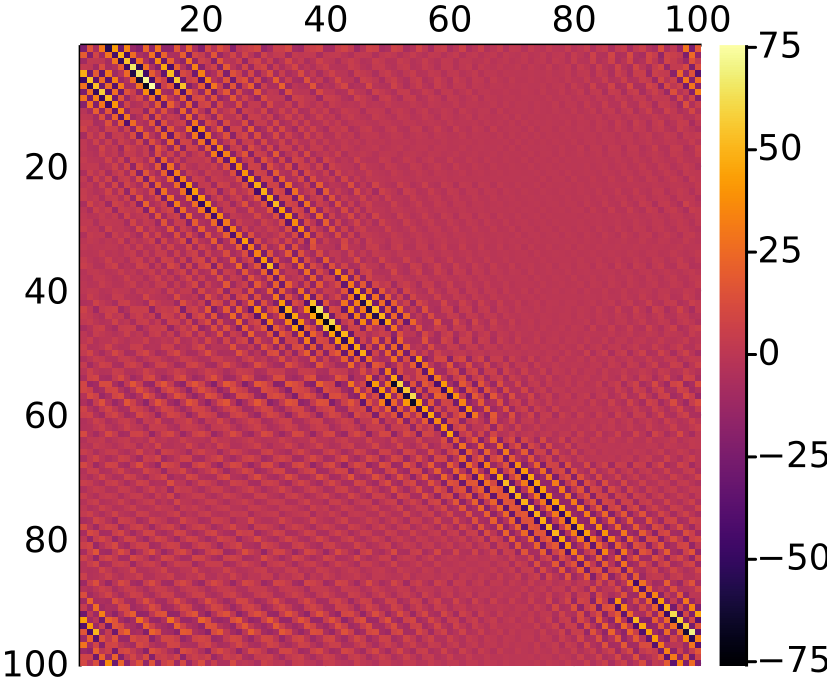} &
            \includegraphics[width=0.22\textwidth]{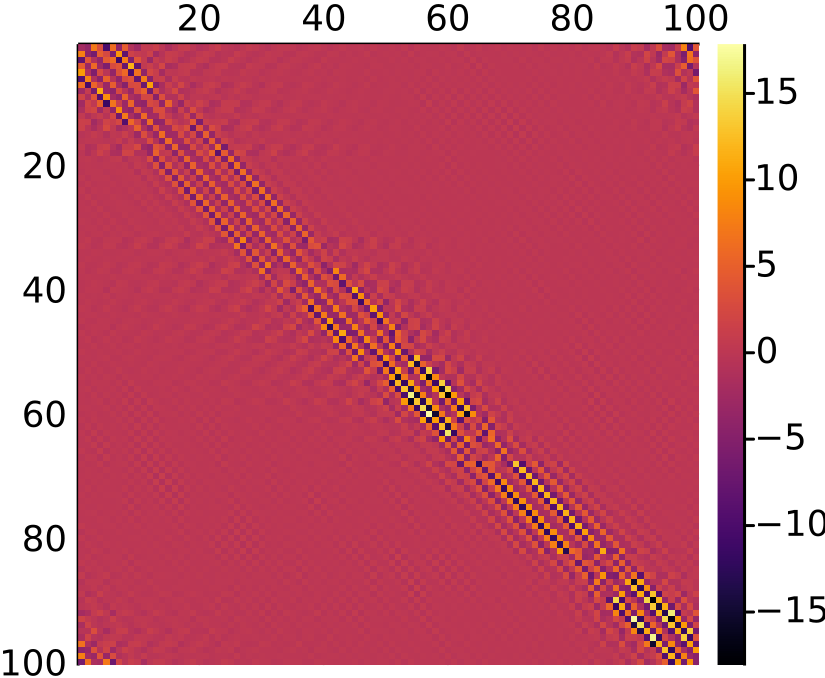} &
            \includegraphics[width=0.22\textwidth]{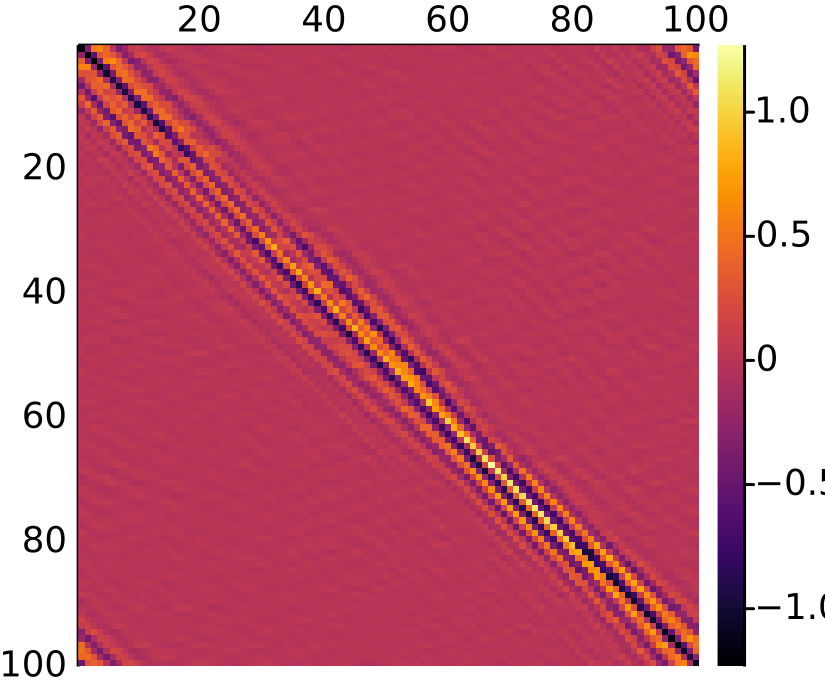} \\
            \rotatebox[origin=l]{90}{Gaussian} &
            \includegraphics[width=0.22\textwidth]{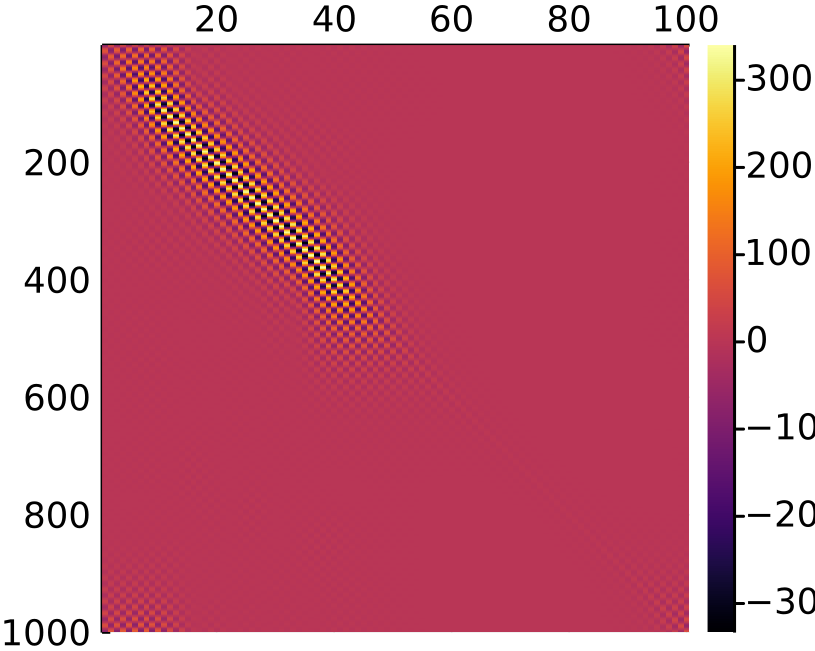} &
            \includegraphics[width=0.22\textwidth]{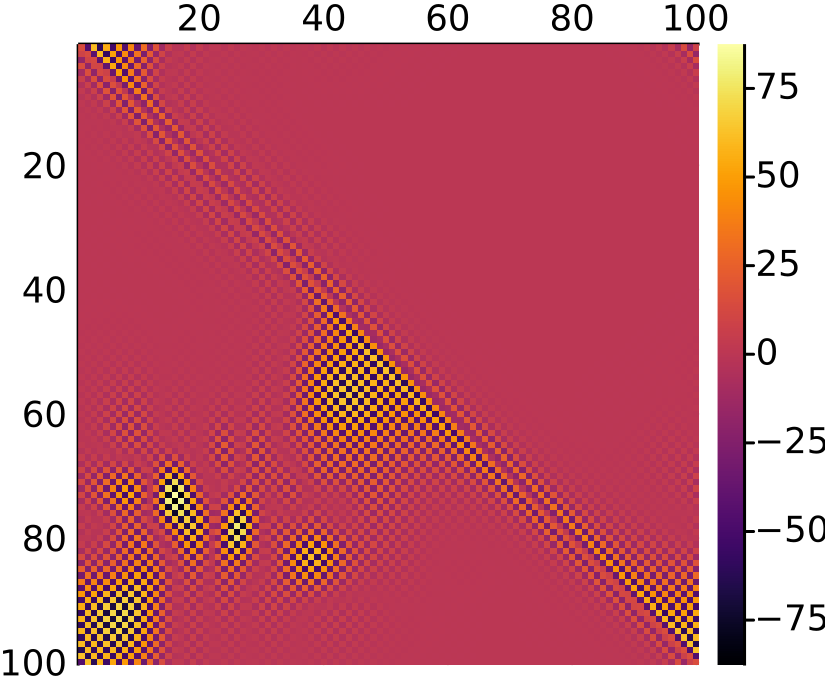} &
            \includegraphics[width=0.22\textwidth]{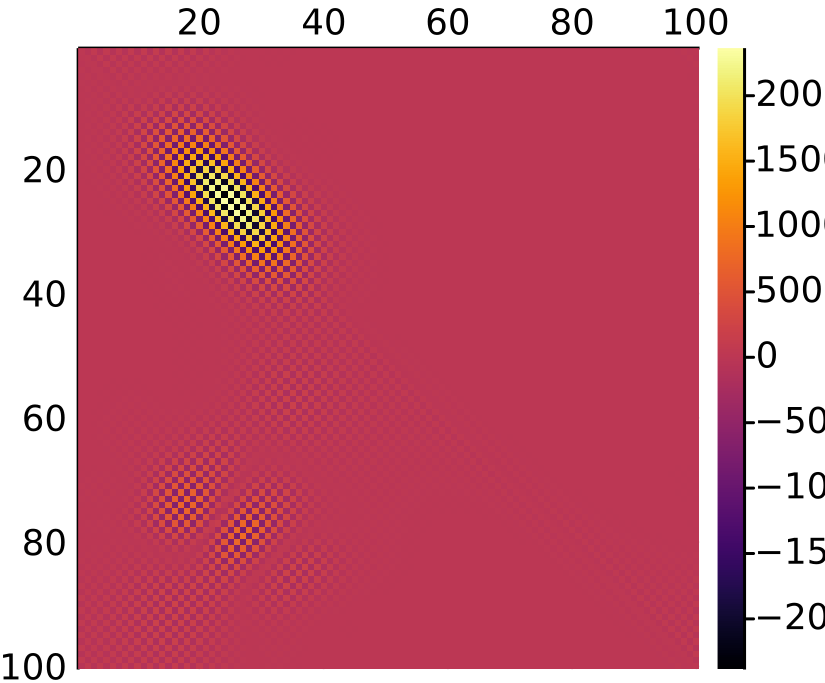} &
            \includegraphics[width=0.22\textwidth]{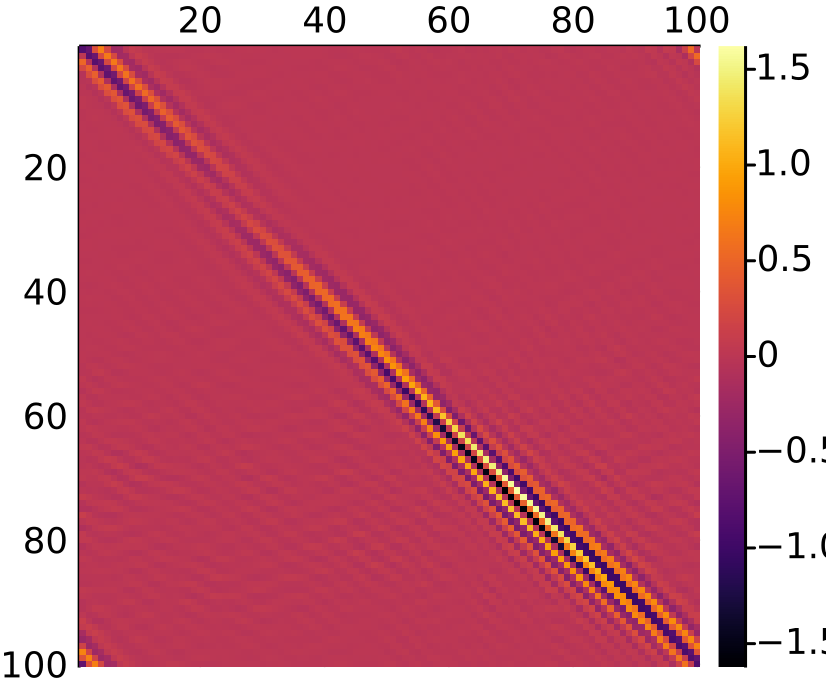} \\
        \end{tabularx}
    \end{center}
    \caption{
        Different operators for $M = 100$. Note that $\mat{R} \in \mathbb{R}^{1000 \times
        100}$ is non-square, despite the appearance. Here, $\mat{A}^{(100)} =
        \operatorname{circ}_{100}\left(\frac{5}{3}, -15, 75, 0, -75, 15,
        -\frac{5}{3}\right)$.
    }
    \label{fig:operators}
\end{figure}

The operators $\mat{R}$ and $\bar{\mat{A}}^\text{DF}$ are both fitted to
$\mathcal{D}^\text{train}$ using least squares (Equations \eqref{eqn:R_data_driven} and
\eqref{eqn:derivative_fit}, respectively), while $\bar{\mat{A}}^\text{emb}$ is fitted using
$10^4$ iterations of stochastic gradient descent with the ADAM optimizer \cite{ADAM} with
step size $0.001$. A grid search over the regularization parameter $\lambda \in \{10^{-12},
10^{-11}, \dots, 10^0\}$ is performed to obtain the smallest time averaged validation error
$\frac{1}{n_t} \sum_{i = 1}^{n_t^\text{valid}}
e_{\bar{\mat{A}}}^\text{valid}(t_i^\text{valid})$, where the error for a given dataset
$\mathcal{D}$ at a time $t$ is given by
\begin{equation}
    e_{\bar{\mat{A}}}^{\mathcal{D}}(t) = \frac{1}{n_\text{IC}^\mathcal{D}}
    \sum_{\bar{\vt{u}} \in \mathcal{D}} \frac{\| \vt{S}(\bar{\mat{A}}, \bar{\vt{u}}(0), t) -
    \bar{\vt{u}}(t) \|}{\| \bar{\vt{u}}(t) \|}.
\end{equation}
The resulting operators are shown in Figure \ref{fig:operators} for both filters
\eqref{eqn:top_hat} and \eqref{eqn:gaussian} with $M = 100$. Note that this grid is
$10\times$ coarser than the DNS grid and cannot represent all frequencies.  While the
top-hat reconstructor and resulting filtered operator are dense, with large off-diagonal
elements, the corresponding Gaussian operators admit a sparser structure. One can clearly
recognize the filter radius profile $h$ in the different operators, with more blurred and
stretched out features on the left half of the domain. Note also that the embedded operator
$\bar{\mat{A}}^\text{emb}$ is closer to the unfiltered operator $\mat{A}^{(M)}$ than
$\bar{\mat{A}}^\text{int}$ and $\bar{\mat{A}}^\text{DF}$, possibly due to a local minimum
near the initial guess $\mat{A}^{(M)}$.

\begin{figure}
    \begin{center}
        \renewcommand{\tabularxcolumn}[1]{m{#1}}
        \begin{tabularx}{\textwidth}{>{\centering\arraybackslash}m{0.02\textwidth}
            >{\centering\arraybackslash}X >{\centering\arraybackslash}X}
            & Top-hat filter & Gaussian filter \\
            \rotatebox[origin=l]{90}{$\frac{1}{n_t} \sum_{t} e_{\bar{\mat{A}}}(t)$} &
            \includegraphics[width=0.45\textwidth]{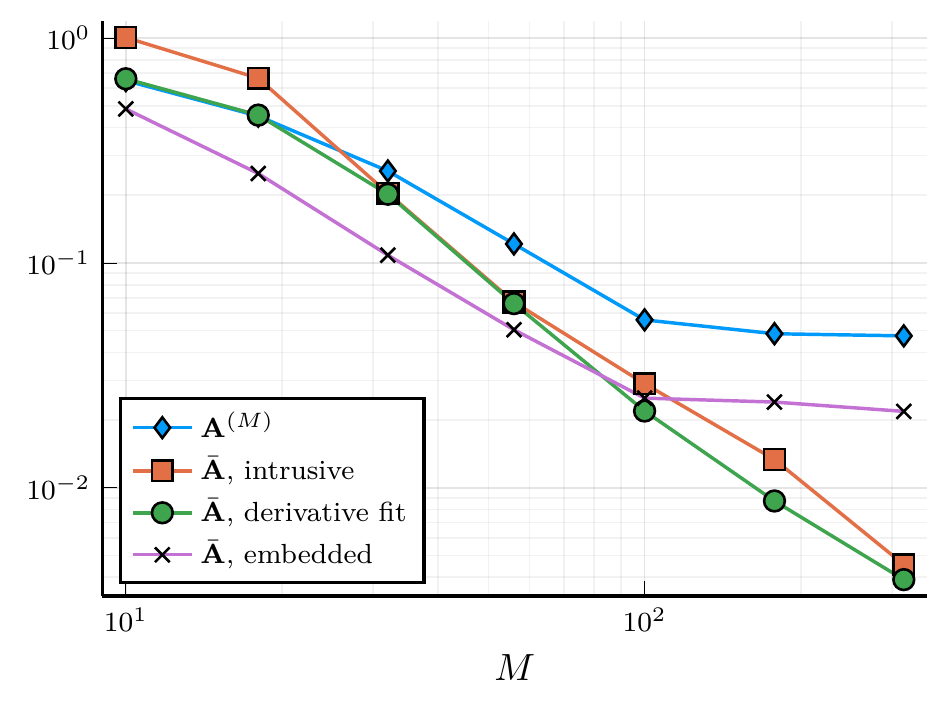} &
            \includegraphics[width=0.45\textwidth]{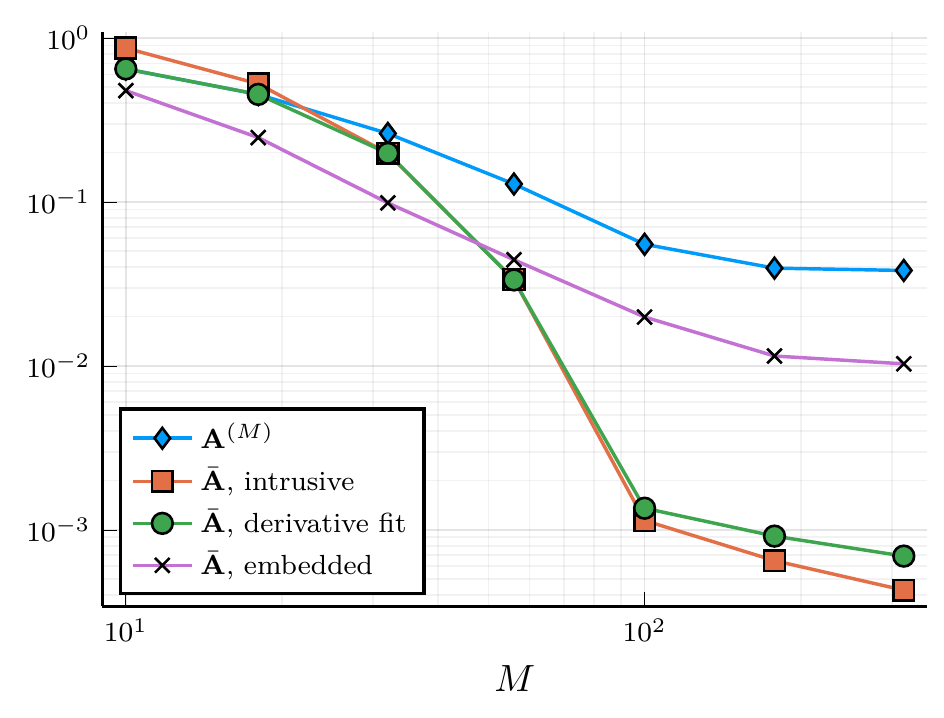} \\
        \end{tabularx}
    \end{center}
    \caption{
        Average relative error on test dataset $\mathcal{D}^\text{test}$ for increasing
        coarse grid precision $M$ for the top-hat filter (left) and Gaussian filter (right).
        The error measures generalization capacity of the ODE-model for new initial
        conditions and time points.
    }
    \label{fig:convergence}
\end{figure}

In Figure \ref{fig:convergence}, the time averaged relative error $\frac{1}{n_t^\text{test}}
\sum_{i = 1}^{n_t^\text{test}} e_{\bar{\mat{A}}}^\text{test}(t_i^\text{test})$ is shown for
both filters as a function of $M$. For the unfiltered operator $\mat{A}^{(M)}$ the error
initially decreases with $M$, but stabilizes at a point between $3\%$ and $5\%$ after $M =
100$. This is due to the true commutator error $\mathcal{C}$, confirming that a closure
model is indeed necessary to predict the filtered dynamics. For $M < 100$ the discretization
error is more important than the commutator error, but this is test-case dependent. The
operators $\bar{\mat{A}}^\text{int}$ and $\bar{\mat{A}}^\text{DF}$ have very similar error
profiles. In fact, they only differ by the regularization terms, as the data fitting terms
are identical (up to a linear transformation). For large $M$, they both admit errors that
are orders of magnitude smaller than $\mat{A}^{(M)}$. Note that for small $M$, the algorithm
gives a high weight to the prior for $\bar{\mat{A}}^\text{DF}$, resulting in the initial
overlap with $\mat{A}^{(M)}$. The embedded operator $\bar{\mat{A}}^\text{emb}$ performs the
best for small $M$, but the errors seems to stagnate like those of its initial guess
$\mat{A}^{(M)}$ for large $M$. A possible reason is that during gradient descent, the
gradients quickly become small, possibly due to vanishing gradients in back-propagation
through the solver. The operator may also have ended up in a local minimum close to the
initial guess. For all the operators, the errors for the Gaussian filters are smaller than
for the top-hat filter, as the higher frequencies are filtered more strongly.

\begin{figure}
    \begin{center}
        \renewcommand{\tabularxcolumn}[1]{m{#1}}
        \begin{tabularx}{\textwidth}{>{\centering\arraybackslash}m{0.02\textwidth}
            >{\centering\arraybackslash}X >{\centering\arraybackslash}X}
            & Top-hat filter & Gaussian filter \\
            \rotatebox[origin=l]{90}{$e_{\bar{\mat{A}}}(t)$} &
            \includegraphics[width=0.45\textwidth]{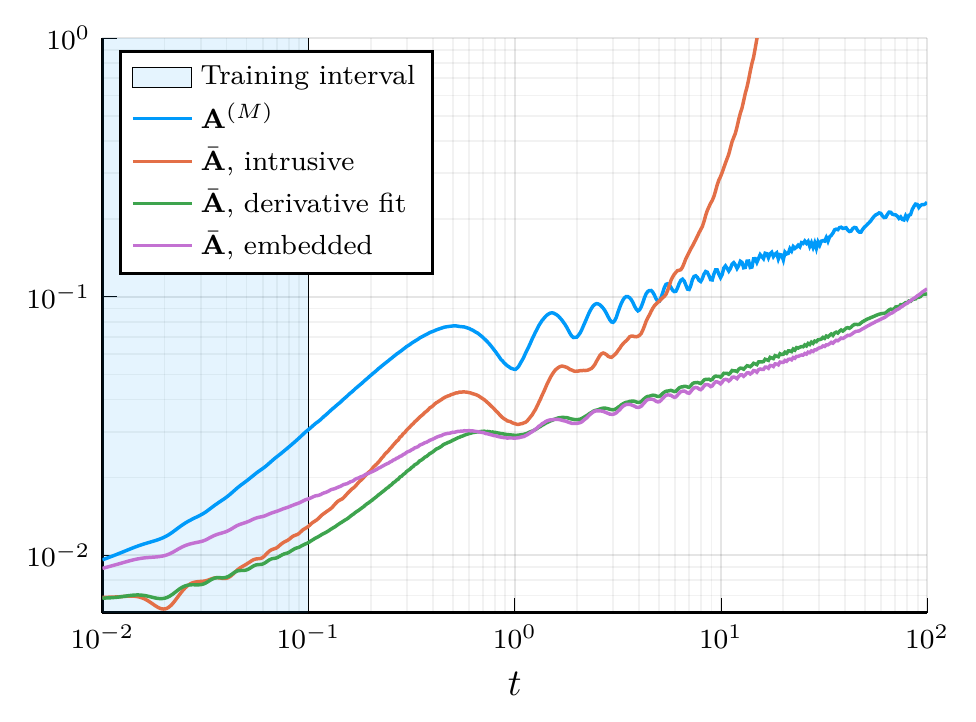} &
            \includegraphics[width=0.45\textwidth]{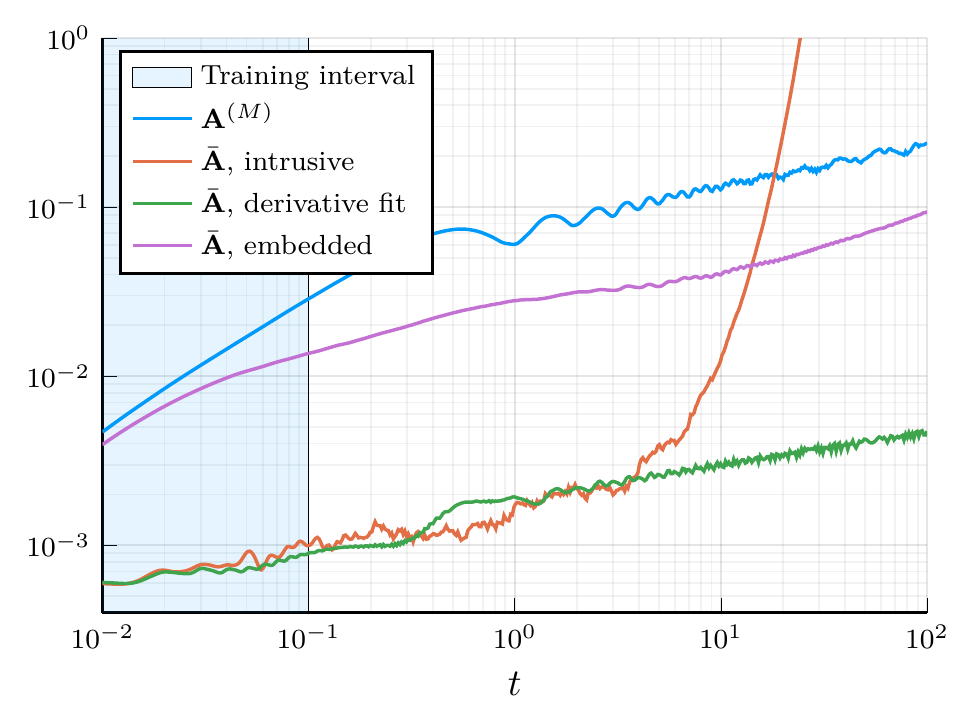} \\
        \end{tabularx}
    \end{center}
    \caption{
        Evolution of the relative error on $\mathcal{D}^\text{long}$ for $M = 100$ using the
        top-hat filter (left) and Gaussian filter (right). The abscissa show the time
        elapsed starting from the filtered initial conditions to one hundred convection
        periods on $\Omega$.
    }
    \label{fig:convection_time_error}
\end{figure}

To further investigate the long term stability of the inferred operators, the evolution of
the mean relative error $e_{\bar{\mat{A}}}^\text{long}$ is computed on the long term testing
data $\mathcal{D}^\text{long}$ for $M = 100$. The results are shown in Figure
\ref{fig:convection_time_error}. At $t = 0$, all operators have an error of zero, since the
initial conditions are given by the filtered DNS solution. The unfiltered coarse-scale
operator $\mat{A}^{(M)}$ initially has the highest errors, as it does not account for the
filter. In addition, the error profile is not fully periodic, but is increasing at a linear
rate. This indicates that the filtered solutions contain frequencies not fully resolved by
the sixth order central difference convection stencil on the coarse grid. The operators
$\bar{\mat{A}}^\text{int}$ and $\bar{\mat{A}}^\text{DF}$ initially perform the best and show
similar error profiles. However, the intrusive operator is not regularized for stability,
and the errors start increasing rapidly after a few convection periods, eventually
surpassing the ones of the unfiltered operator. This is likely due to instabilities from
anti-diffusion on the parts of the domain where the filter radius $h$ is decreasing (on
$\left[\frac{1}{4}, \frac{3}{4}\right]$). For $\bar{\mat{A}}^\text{DF}$ the errors seem to
be increasing at the same rate as $\mat{A}^{(M)}$. The errors for the embedded operator
$\bar{\mat{A}}^\text{emb}$ stay about one order of magnitude below those of $\mat{A}^{(M)}$,
and increase at the same rate.

Note that putting the continuously filtered initial conditions $\bar{u}_0$ into the
unfiltered continuous equations \eqref{eqn:continuous} would lead to errors as the
commutator error $\mathcal{C}$ would be not be accounted for. But the error profile would
still be periodic, as the exact solution to \eqref{eqn:continuous} is periodic for all
initial conditions. In particular, the error would be exactly zero when $t = n$ for all $n
\in \mathbb{N}$. This does not seem to be the case for the unfiltered reference operator
$\mat{A}^{(M)}$. The inferred operators are thus not only learning to account for the
explicit filter $\mathcal{F}$, but also for the under-resolved discretization
$\vt{x}^{(M)}$.

\section{CONCLUSION} \label{sec:conclusion}

In this article we have considered the effect of applying discrete non-uniform filters to a
linear partial differential equation. Different discrete closure models were considered,
using filtered data samples to infer operators. Firstly, we built an explicit reconstruction
model which allows us to use the DNS equations to evolve the system in time. This approach
gave satisfactory short-term errors, but was shown to lack long-term stability. It also
requires having access to the full order model operator $\mat{A}$ on the fine
discretization. Secondly, we built the discrete filtered operator directly, using
gradient-free and embedded approaches. The former gave similar error profiles to the
intrusive approach, while the latter gave less significant improvements for the cases
considered. When building the filtered operator directly, we learn not only the correction
for the filter, but also possibly improving the finite difference approximations to the
unfiltered operator itself.

In the case of the convection equation, the least-squares time-derivative fit on the coarse
grid was found to be just as performant as the one involving the DNS-grid, being sufficient
for inferring operators with good extrapolation and generalization capacities. On the other
hand, the embedded model did not achieve the same performance. It requires the choice of
more hyperparameters, such as the initial guess, number of iterations, step size,
regularization, and choice of sensitivity algorithm for computing the gradients. Care must
be taken to avoid local minima.

In conclusion, this study shows that constructing a discrete closure model with an embedded
strategy, even for a linear model problem, leads to a complicated optimization problem with
many choices (such as hyperparameters) and does not easily lead to more accurate or stable
results than the more straightforward `derivative fitting' approach. This indicates that
more advanced parameterizations, such as neural networks, are also expected to require very
careful consideration in order to work in an embedded set-up.

\section*{Software availability statement}

The source code for the simulations and figures is available at
\url{https://github.com/agdestein/DiscreteFiltering.jl}. All the simulations were run on a
recent laptop computer.

\bibliographystyle{abbrv}
% \bibliography{library/references}

\end{document}